\documentclass{iaus}
\usepackage{graphicx}

\title[A new look at NICMOS transmission spectroscopy] 
{A new look at NICMOS transmission spectroscopy: no conclusive evidence for molecular features}

\author[N. P. Gibson et al.]{
N. P. Gibson$^{1}$,
F. Pont$^{2}$ \and
S. Aigrain$^{1}$}

\affiliation{
$^1$Department of Physics, University of Oxford 
$^2$School of Physics, University of Exeter
}

\pubyear{2010}
\volume{276}  
\jname{The Astrophysics of Planetary Systems: Formation, Structure, and Dynamical Evolution}
\editors{A. Sozzetti, M. G. Lattanzi \& A. P. Boss, eds.}
\begin{document}

\maketitle

\begin{abstract}
We present a re-analysis of archival HST/NICMOS transmission spectroscopy of the exoplanet system, HD 189733, from which detections of several molecules have been claimed. As expected, we can replicate the transmission spectrum previously published when we use an identical model for the systematic effects, although the uncertainties are larger as we use a residual permutation algorithm in an effort to account for instrumental systematics. We also find that the transmission spectrum is considerably altered when slightly changing the instrument model, and conclude that the NICMOS transmission spectrum is too dependent on the method used to remove systematics to be considered a robust detection of molecular species, given that there is no physical reason to believe that the baseline flux should be modelled as a linear function of any chosen set of parameters.
\keywords{stars: individual (HD 189733), planetary systems, techniques: spectroscopic
}
\end{abstract}


Transmission spectroscopy is a powerful technique that can probe the atmospheres of transiting planets for atomic or molecular species \cite[(e.g. Seager \& Sasselov 2000; Brown 2001)]{Seager_2000,Brown_2001}, by measuring the wavelength dependence of a planet's observed radius. The size of a planet is determined by the altitude at which the atmosphere becomes opaque to starlight, which may vary due to atomic and molecular absorption.

HST/NICMOS transmission spectroscopy has led to some of the most detailed studies of exoplanet systems to date. However, instrumental systematics are larger than the expected signal due to molecular absorption, and consequently the methods used to remove the instrumental systematics have a considerable effect on the output transmission spectra. \cite[{Swain}, {Vasisht}, \& {Tinetti} (2008, hereafter SVT08)]{Swain_2008} presented a transmission spectrum of HD 189733, claiming detections of H$_2$O and CH$_4$. We present a detailed re-analysis of this and other NICMOS data sets in \cite[{Gibson}, {Pont}, \& {Aigrain} (2010, hereafter GPA10)]{GPA_2010}, and argue that the detection of molecular species is dependent on the choice of instrument model, and therefore cannot be considered as robust. We only briefly summarise our findings here, and refer the reader to GPA10 for further details.

After extracting the raw light curves for each wavelength channel, we first model the systematics using an identical instrument model to that of SVT08. Not surprisingly, we produce a very similar transmission spectrum, although with some disagreement at the blue end and with larger uncertainties. The larger uncertainties are the result of using a residual permutation algorithm, which attempts to take the remaining systematic noise and correlations between the instrument model parameters into account. If we instead evaluate the transit depth using a Levenberg--Marquardt algorithm, we get almost identical uncertainties to those reported in SVT08. However, this method does not properly take all sources of uncertainty into account. We further note that the residual permutation algorithm does not fully account for all sources of uncertainty, in particular those that arise from offsets in flux level between the HST orbits not properly corrected for by the instrument model. Offsets in the in-transit orbit are fitted for by the transit model, and do not appear in the residuals to contribute to the measured uncertainties. Hence the `true' uncertainties are likely even larger than we reported.

We also produce transmission spectra of HD 189733 using slightly different instrument models to remove the systematics; one using extra quadratic terms in the decorrelation function, another using only two of the three out-of-transit orbits to fit for the instrument model, and another excluding the angle parameter from the instrument model (see GPA10 for details). Using different instrument models we produce rather different transmission spectra. Given that there is no physical reason to assume the instrument should follow a linear function of a chosen set of optical state parameters, there is no reason to prefer one model over another. We therefore argue that further physical justification for the specific instrument model used in SVT08 (and for many other important NICMOS results) is required before these spectra can be considered as robust.

Recently \cite[{{Deroo}, {Swain}, \& {Vasisht} (2010, hereafter DSV10)}]{DSV10} posted a response to our paper, claiming that our analyses were flawed. They argue that the larger uncertainties we produce in our transmission spectrum are the result of a noisy instrument model, and that the subsequent analyses using varying instrument models are therefore invalid. As stated earlier, the larger uncertainties are due to a different method used to evaluate them. In GPA10, we carefully compared the instrumental parameters used to model the systematics to those provided in SVT10 (supplementary material), and concluded that they show a similar dispersion and amplitude; so why do the plots in DSV10 appear to contradict this? {\it The answer is that the data marked ``Swain et al. (2008)'' in DSV10 are not the same data as shown in SVT08} \footnote{As an appendix to the astro-ph posting of this article, we provide plots comparing the SVT08, GPA10, and DSV10 datasets.}. The plots in DSV10 are therefore misleading, and their conclusions are based on a misrepresentation of data.

DSV10 also criticises our treatment of the XO-1 data from \cite{Tinetti_2010}, stating that the reason we cannot reproduce the same transmission spectrum is because we omit some parameters from the decorrelation. This is simply re-stating what we already explained in GPA10. The parameters were omitted from the instrument model as these would require extrapolation to the in-transit orbits. Unfortunately, almost no detail is given in \cite{Tinetti_2010} regarding the data analysis, and we could not compare instrumental parameters and identify the source of the discrepancy.

One of the primary goals of our paper is to encourage open discussion within the exoplanet community about the reliability of methods used to remove systematic noise from this type of dataset, and the robustness of derived results. In our opinion this is a very challenging, unsolved problem. We strenuously contest the claim made by DSV10 that our paper is a confirmation of their results, but will continue to seek a better understanding of the discrepancies between our results and theirs. We are also actively testing new methodologies for the robust characterisation of transmission spectra in the presence of strong systematics, and are in the process of analysing STIS and WFC3 observations of this object, to get a complete transmission spectrum from UV to NIR.

\newpage

\begin{figure*}
\begin{center}
 \includegraphics[width=125mm]{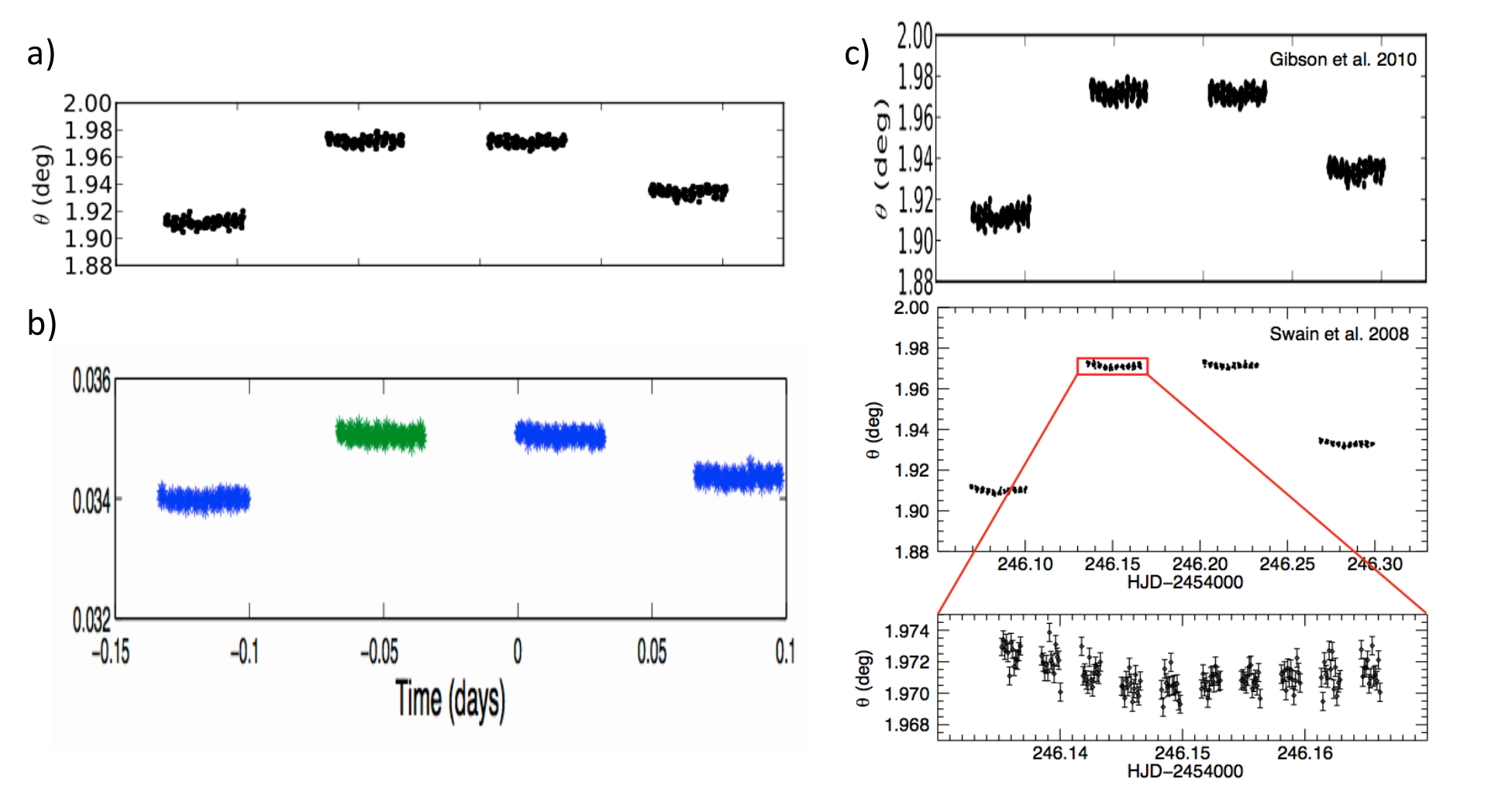} 
\caption{The central argument of DSV10 is that the instrument model of GPA10 is noisier than that of SVT08. They illustrate this using the angle of the spectral trace on the detector, which is one of the parameters of the instrument model. Here we reproduce the relevant plots from the original papers: a) GPA10, b) SVT08, c) DSV10. DSV10 label their plot as showing data from SVT08, but the relative scatter of the data in b) and c) (middle panel) are clearly different. However, a direct comparison is made difficult by the different units and y-axis scales of the three plots.}
   \label{fig1}
\end{center}
\end{figure*}

\begin{figure*}
\begin{center}
 \includegraphics[width=125mm]{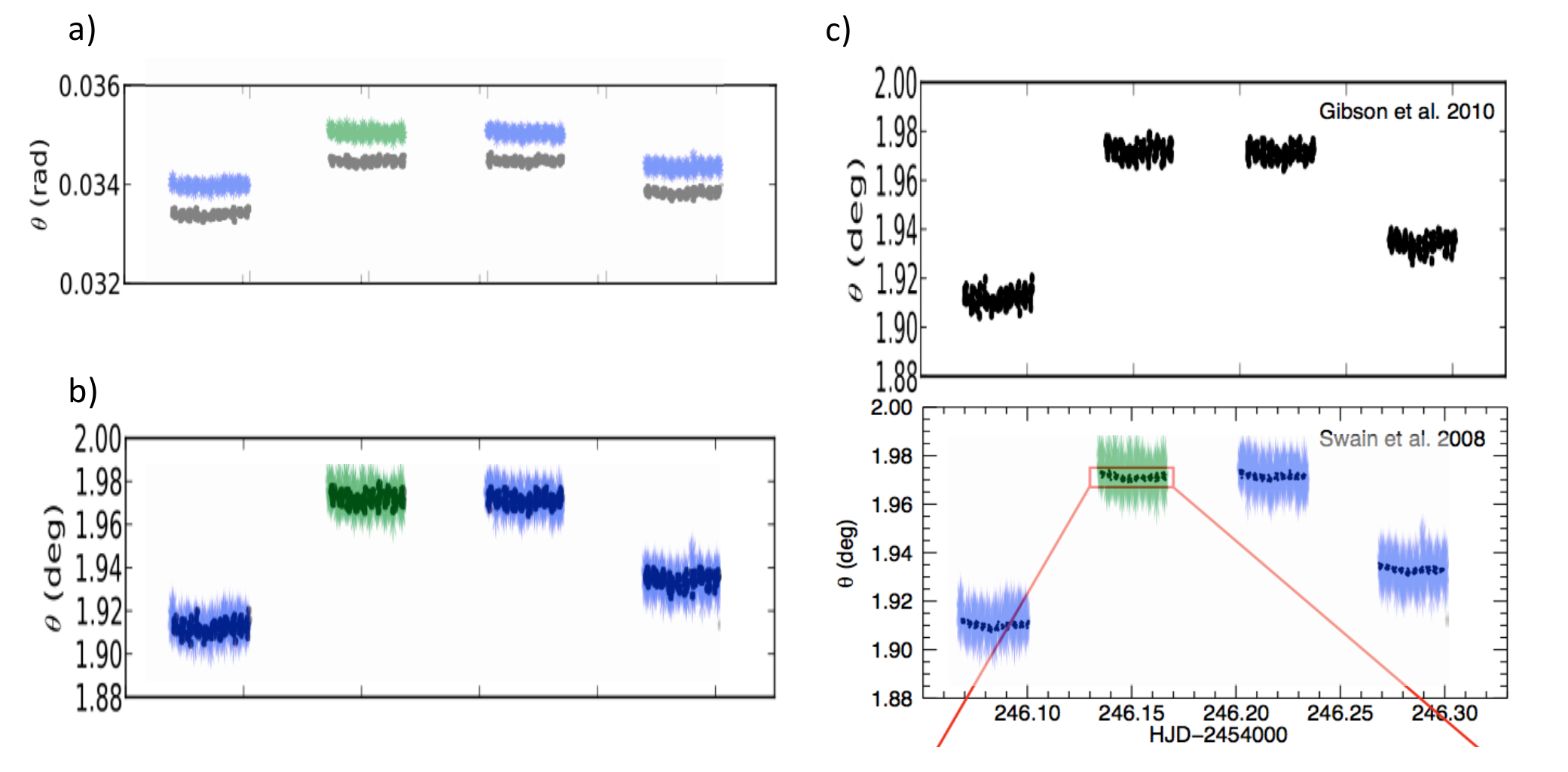} 
\caption{We now compare the datasets from GPA10, SVT08 and DSV10 after re-plotting to the same scales and translating and stretching the above plots. a) shows the data from GPA10 and SVT08 plotted on the same y-axis (radians) scale. The noise level of the GPA10 and SVT08 datasets are similar (it is difficult to make a detailed comparison owing to the relatively large size of the points used by SVT08), although there is a (real) zero-point offset. The dataset shown in DSV10 (yet attributed to SVT08) has a much lower scatter than both of the former (and the same zero-point as GPA10, i.e. different from the SVT08 zero-point). In other words, the data presented in DSV10 are not the data used by SVT08, and the claim that the GPA10 instrument model is much noisier than the SVT08 instrument model is incorrect. To further illustrate this, b) shows the data from SVT10 scaled and translated to overlay the GPA10 data, and c) shows the SVT10 data scaled and translated to overlay the DSV10 plot (now removing the offsets).
}

\label{fig1}
\end{center}
\end{figure*}

\end{document}